# Vehicle Route Prediction through Multiple Sensors Data Fusion

**Ali Nawaz**

Department of Computer Engineering,
College of Electrical Mechanical Engineering (CEME),
National University of Science and Technology (NUST),
Islamabad,
Pakistan
anawaz.cse19ceme@ce.ceme.edu.pk

**Attique Ur Rehman**

Department of Computer Engineering,
College of Electrical Mechanical Engineering (CEME),
National University of Science and Technology (NUST),
Islamabad,
Pakistan
aurrehman.cse19ceme@ce.ceme.edu.pk

**Abstract**

Vehicle route prediction is one of the significant tasks in vehicles mobility. It is one of the means to reduce the accidents and increase comfort in human life. The task of route prediction becomes simpler with the development of certain machine learning and deep learning libraries. Meanwhile, the security and privacy issues are always lying in the vehicle communication as well as in route prediction. Therefore, we proposed a framework which will reduce these issues in vehicle communication and predict the route of vehicles in crossroads. Specifically, our proposed framework consists of two modules and both are working in sequence. The first module of our framework using a deep learning for recognizing the vehicle license plate number. Then, the second module using supervised learning algorithm of machine learning for predicting the route of the vehicle by using velocity difference and previous mobility patterns as the features of machine learning algorithm. Experiment results shows that accuracy of our framework.

Keywords:    Machine learning, Deep learning, Supervised learning

## 1. Introduction

The aim of artificial intelligence (AI) is to make the human life more comfortable and reduce the effort of human to perform certain tasks. Vehicle driving is one among those tasks where human put maximum effort to reach his destination. From last two decades, scientist researching in this field and have achieved good results but still there are some risks and issues involved. Traditionally, there are two types of research aspects in this field first is making autonomous or driver less vehicle [1] and second is making complete network among the vehicles [2] where vehicles share their routes knowledge among them. When we talk about the autonomous vehicle there are lot of problems associated with it. The foremost problem with autonomous vehicle is privacy [3] as autonomous vehicle requires the connection between your own vehicle and personal devices like smart phone that is how it will be able to manage your schedule and make intelligent decisions now the privacy question rise here is that autonomous vehicle might store your daily route information, your daily schedule and location you visits someone might steal these information for achieving wrong goals or companies purchase your information to provide advertisement of their products. The second and most important problem in autonomous vehicle is hacking [4] as autonomous vehicle require continuous internet connection to communicate and for path detection and any devices that are connected via internet is always at risk of being hacked someone might hack the information of autonomous vehicle and provide wrong route information which will disturb the communication between vehicle and cause an accidents also autonomous vehicle require your personnel bank account information like bank account number and ATM pin etc. someone might hack these information and make you bankrupt.

Privacy is the major issue in vehicle adhoc network (VANET) [5] as in VANET vehicles shares their routes knowledge among them someone might use these knowledge to achieve his bad objectives. Security is also the important concern in VANET someone might hack the networking between vehicles and provide wrong route information which will be the cause of an accidents [6]. The vehicle driving is always a risky task either you are holding the steering wheel or walking on your feet on the road the chance to meet unfortunate accidents are always there. If there are some unfortunate accident took place then who is responsible for it whether autonomous vehicle itself, manufacturer of autonomous vehicles or internet distributed companies? [7] That is why, we cant simply ignore the importance of humans in vehicle driving.

When the vehicles reach the crossroad, it is difficult for cross vehicles to understand that whether opposite vehicle is going to change its route or not i.e. either it is going to take turn or going straight. For example, C1 is one vehicle and is travelling on the right side of road as well as travelling along the one side of crossroad and C3 is another vehicle and travelling along the left side of road which is on the other side of crossroad now it is difficult for C1 to understand whether the C3 is going to take turn or travelling in same route and vice versa.

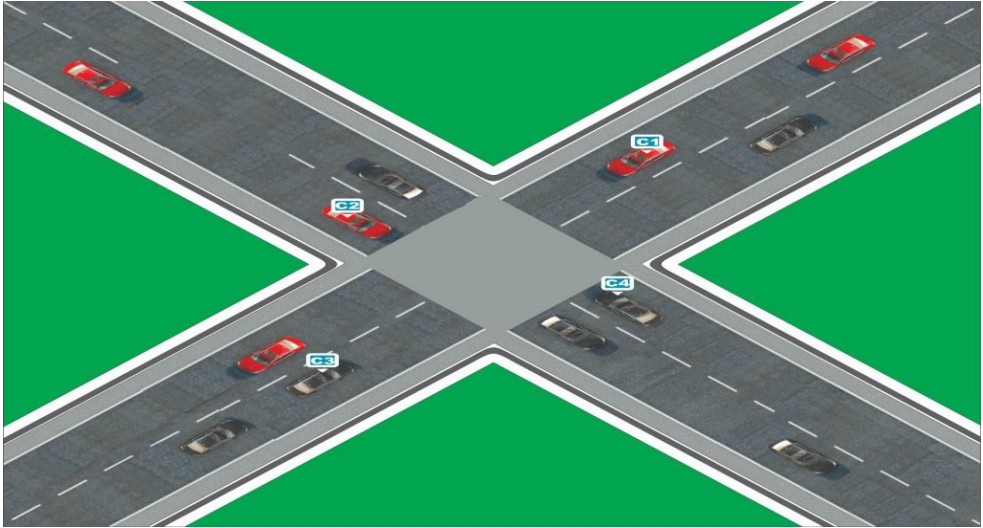

**Figure 1.** Illustration of the problem statement.

These are some major issues associated with route prediction in autonomous vehicles and vehicle communication. Now, in order to overcome these issues in route prediction we proposed a system which will combine both machine learning [8] for making prediction on the basis of certain features and deep learning for detecting the license plate number of vehicles. With the development of advance sensors such as digital camera and deep learning [9] libraries like Keras and TensorFlow nowadays our problems of computer vision are become simpler with more accuracy that is why we use deep learning for detecting and recognizing the license plate number of target vehicles. Then we use machine learning technique to predict whether vehicle travels straight path or make turn in crossroad. In order to achieve this goal of prediction we will use radar gun for velocity detection and global positioning system (GPS) for detection of previous mobility patterns.

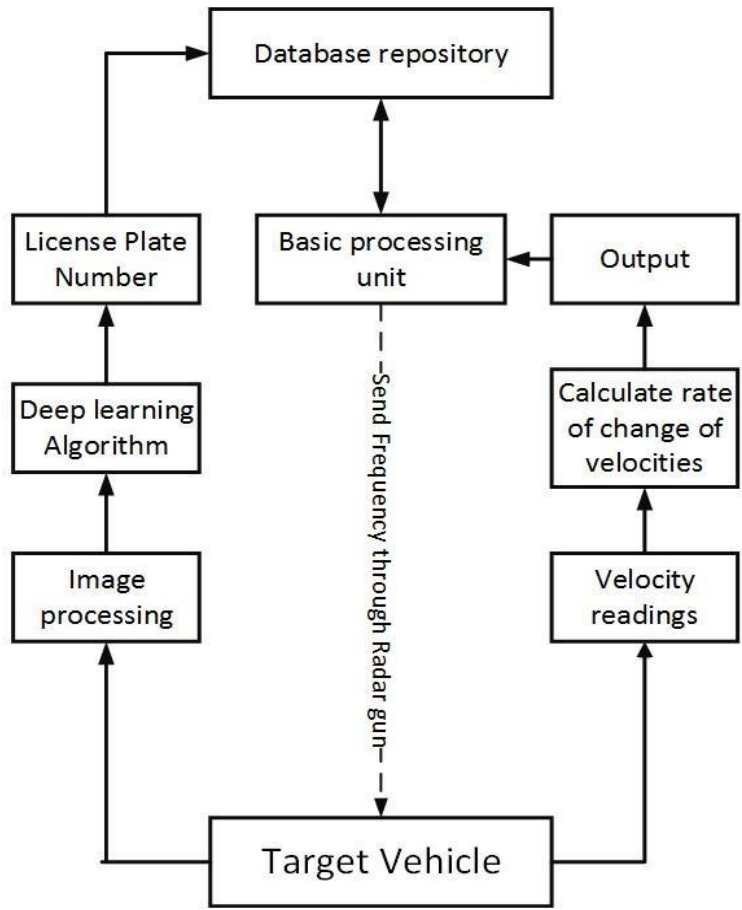

**Figure 2.** Architecture of proposed system

Our proposed system detects vehicle license plate number by using camera and recognize it by using deep learning technique after recognizing our proposed system check for license number in database repository if database repository does not contain the database of specified vehicle then terminate the remaining process otherwise send the control to the basic processing unit which will send frequency to the target vehicle in order to measure velocities afterward output of velocities difference is send to the basic processing unit (B.P.U) to predict route of the vehicle.

Because of aforementioned limitation of autonomous vehicle and VANET we proposed a system which is neither fully autonomous nor fully control by humans our proposed system will involve both human intelligence (H.I) and A.I in order to maximize the accuracy of vehicle driving as well as route prediction which in turn results in reducing the probability of the accidents. H.I + A.I = More Accuracy in driving

The major contributions of our research paper are summarized as follows.

- We have proposed a method for measuring the difference of speeds and taken it as a feature for machine learning algorithm

- We have proposed efficient system for crossroad route prediction

- Reduce the probability of the accidents and overcome the issues of autonomous vehicles and VANET

The rest of the paper is divided into four sections which are organized as follows. Section II will present related work. In section III we will describe our proposed method in details and Section IV contain our experiment. The V and final Section concludes our research paper.

## 2. Related work

When we look into the literature of vehicle route prediction there are two most commonly used techniques for predicting the routes first is by analyzing the driver intentions and previous behavior while second is by making the future trajectory of the targeted vehicles. Both techniques has their own pros and cons. We will review both techniques with their known limitation. In [10], proposed a statistical model for predicting the route of vehicle based on hidden Markov model (HMM)[11]. The probabilistic model predicts the intent of driver by previous route history data of the route followed by vehicle driver obtained via GPS. The basic idea is to assemble data on each route a driver takes and to invigorate the truthful model based on that experience. The model is then used, on-line, to anticipate driver plan for the accompanying route, after which that trip is used to, eventually, update the model. Instead of predicting the route the model put focus on the predicting the behavior of the vehicle driver.

[12] present an algorithm for turn prediction of the vehicle drivers. This algorithm is simply based on the fact that driver likely to take turn which take short time and resources to his destination. The accuracy of the proposed algorithm increases by using more representation of destination.

In 2008, an algorithm for predicting the short-term route of vehicles drivers by means of Markov model [13]. In this algorithm he actually makes the prediction by training the model on previous long-term route history data of vehicles drivers obtained by GPS. The proposed algorithm divides route into n segments where n ranges from 1 to 10 and predict 90 percent accuracy for first individual segment and accuracy consistently decreases with each next number of segments. The only limitation of this algorithm is the distance as each next segment prediction is limited to 237.5 meters.

In the same year,[14] developed an algorithm in which they eliminate the segmentation limitation and predict the route of the vehicles based on the previous route history obtained by GPS data. The proposed algorithm made the prediction by comparing the current trip of vehicle by previous route knowledge of the same vehicle. The only feature use in this algorithm is previous route history obtained by GPS data due to which accuracy of the prediction varies and, in some cases, decreases up to 50 percent.

In [15], proposed Variable-order Markov Models (VMMs) for analyzing and predicting the short-term route prediction based on the data collected from Shanghai taxis. The prediction accuracy of proposed method increases with the increasing in available data. The proposed approach is centralized which results in bad scalability when number of vehicles increases.

In [16], proposed a probabilistic trajectory prediction approach based on Gaussian mixture model. The proposed approach achieve this task in two seconds advance by concluding the joint probability distribution obtained by previous mobility patterns. The idea behind the approach is to gain motion model from the recently observed trajectories and make prediction of the future trajectories. The proposed method not only make prediction but also make statistical distribution of the future trajectories. Although the proposed approach is trained on 1200 trajectories and evaluated on 21 test trajectories the accuracy is pretty efficient but the only limitation is the future prediction time two seconds which is not appropriate for decision making.

In [17], proposed the method for the predicting the trajectory of the vehicle. The proposed method predicts the new trajectory by combining the prediction of trajectory obtained via Constant Yaw rate acceleration motion model (CYRA) and trajectory of prediction obtained by maneuver recognition. The maneuver is totally related to the route of the vehicle. The maneuver recognition model (MRM) depends on early detection of the path where the driver is expecting to go. The lane objective is to detect and exploits the present resemblance between the method for the vehicle under consideration and the point of convergence of the lane. The CYRA model is mathematical model which is obtained by integrating the velocities. The preprocessing is obtained by performing a Kalman filter with a CYRA motion model for the purpose of prediction. Although proposed method works on combination of multiple trajectories, the accuracy of proposed method is bad for longer term prediction and non-monotonic movement as well.

Similarly, [18] propose a method to detect the change of lane trajectory by gathering and analyzing different driver's data. The proposed method produces good accuracy of detecting any usual driver lane change behavior. The only problem of this approach is the limitation of trajectories set used as compared to the traditional trajectory approach.

There are two ways of detecting the object; first is discriminative and other is generative methods [19]. In discriminative, first we detect all object in image or frames of the video and after that re-distinguish them to extend a follow associating a similar object. The generative method generates the complete model of the object with the goal of making correlation between the images and the model, which enhances the exactness of the matching. We will use discriminative approach to detect license plate of the vehicles which is character as well as object detection and recognition task. The most prominent approach of route prediction is traditional vehicle route prediction used by autonomous vehicle [20] in which vehicle predict others vehicle future trajectories but due to early mentioned cons of autonomous vehicle we should not completely relying on this technique. Another method to predict the route of the vehicles is using the navigation information [21].

**Table 1.** Route prediction techniques

| Source | Proposed Method | Advantage | Disadvantage |
| --- | --- | --- | --- |
| [10] | Statistical model for predicting the route of vehicle based on hidden Markov model (HMM) | Anticipate driver plan for the accompanying route | Focus on predicting the behavior of the vehicle driver |
| [12] | Take the turn on the basis of driver shortest route to his destination | Accuracy increases with more representation of destination | |
| [13] | Predicting the short-term route of vehicles drivers by means of Markov model | Predict 90 percent accuracy for first individual segment | Each next segment prediction is limited to 237.5 meters |
| [14] | Predict the route of the vehicles based on the previous route history obtained by GPS data | Eliminate the segmentation limitation of [12] | Accuracy of the prediction decreases up to 50 percent |
| [15] | VMMs for analyzing and predicting the short-term route prediction | Accuracy increases with the increasing in available data | Poor scalability |
| [16] | Probabilistic trajectory prediction approach based on Gaussian mixture model | Not only make prediction but also make statistical distribution of the future trajectories | Future prediction time is two seconds which is not appropriate for decision making |
| [17] | Predict trajectory by combining the trajectory obtained via (CYRA) and trajectory of prediction obtained by maneuver recognition | Work on combination of multiple trajectories | The accuracy of proposed method is bad for longer term prediction and non-monotonic movement as well |
| [18] | Method of detecting the change of lane trajectory | Good accuracy on few trajectory set | Doesn't work well for unusual driver behavior |
| [20] | Vehicle predict other vehicles future trajectories | | Privacy and Security is the major risk |
| [21] | Classify mobility based on the route prediction in VANET | The proposed scheme efficiently reduces handoff latency | Hacking and Privacy is the major risk involved in VANET |

## 3. The proposed method

We will discuss our proposed method in detail later in this section but first let's discuss the purpose and working of B.P.U.

*3.1.* Basic processing unit

Our basic processing unit shown in fig.2 is composed of three sub modules which are responsible for following mentioned tasks.

- Send frequency by radar gun for measuring velocity

- Machine learning algorithm for predicting route

- Display route prediction result to driver in order to make decision

The internal architecture of B.P.U is show below:

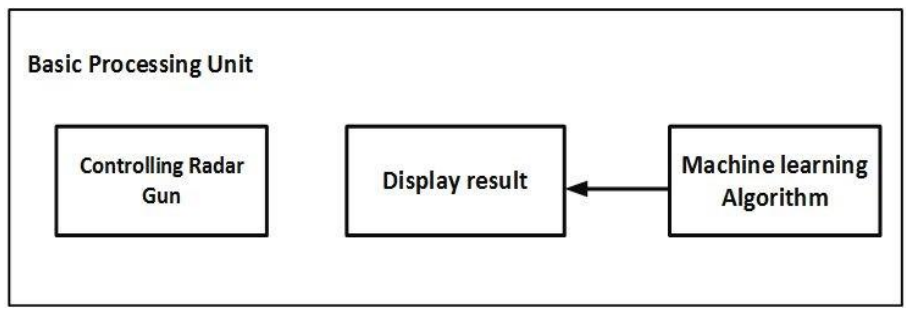

**Figure 3.** Submodules of B.P.U

Left hand side module is responsible for controlling the operation of radar gun for velocity measure. Right hand side module is responsible for predicting route based on machine learning technique and central module is responsible for displaying the prediction result.

*3.2.* Working of our proposed system

Our proposed method consists of two different modules but work together in a sequence. One module is for recognition of vehicle by its license plate or registration number and other is for predicting route based on machine learning technique. Our proposed system begins by detecting the license plate number of target vehicle by using camera and then recognize the plate number by using deep learning via tesseract. After recognizing the vehicle license plate number, it will check whether corresponding vehicle license number is registered in our database system or not. If it is registered in our database, then send the control to B.P.U otherwise terminate the whole prediction process. In B.P.U, radar gun control procedure sends some initial frequency and get first velocity reading v1 by using the formula
v1= $\Delta f1$ / fo1 where $\Delta f1$ = fr1 − fo1

fr1 = first reflected frequency from target fo1 = first initial frequency we send from radar gun and after 5 sec later send another frequency beam and get second velocity reading v2

v2 = = Δf2 / fo2 where Δf2 = fr2 − fo2

fr2 = second reflected frequency from target fo2 = second initial frequency we send from radar gun after getting the value of v1 and v2, calculate the rate of change of veloc-

ities by using the equation:

Δv = v2 − v1

The rate of change of velocity is taken as one of the features of machine learning for the task of route prediction as our task is supervised learning task in which we have features and labels. So, in this task our datasets are composed of two features; rate of change of velocity and previous mobility pattern data which are also stored in database repository. After getting rate of change of velocity Δv, we apply machine learning algorithm which will take rate of change of velocity and previous mobility patterns as features and label them with S for straight and T for turn depends on the features. At last the result is displayed on screen and driver will decide whether to go straight or to take turn on the basis of prediction made by the system.

### 3.3. Machine learning techniques

As We mentioned earlier that our goal is to classify whether opposite vehicle will going to change its route or not it means that our task is classification task of supervised machine learning. There are numerous machine learning algorithms which deals with classification problems. In order to solve machine problem, we are going to use most prominent and simple algorithms of classification machine learning which are explained below

### 3.3.1. K-nearest neighbor classifier

K-nearest neighbor is the simple and prominent machine learning algorithm widely used for classification problem of supervised learning. It finds the distance between test point with every point in the training data then find the k nearest neighbor for that training point [22].

### 3.3.2. Naive Bayes Classifier

Naive Bayes is another most prominent, simple and widely used machine learning for solving classification task. It is based on the concept of conditional probability to classify the test data into predefined classes. Conditional probability is the probability that something will happens depends on conditions. let's consider two events A and B, then the conditional probability is defined as

$$P(A \text{ and } B) = P(A)*P(B \mid A)$$

Naive Bayes classifier locate the individual likelihood of each element and then select the component with highest likelihood.

*3.3.3.* Decision Tree Classifier

Decision tree is the class of supervised machine learning model that can be used both for classification and regression. In decision tree classification, we predict that every anticipation belong to the most usually occurring class of training perceptions in the area to which it belongs [23].

*3.4.* Datasets

Our machine learning datasets would be like this:

**Table 2.** Sample Dataset

| Velocity Difference Δv | Mobility patterns | Prediction |
|---|---|---|
| −3 | 1 | Turn |
| 0.1 | 0 | Straight |

It shows our datasets consist of two features one is velocities difference ΔV and other is previous mobility patterns M.P. Prediction is the label of our datasets which might be straight which means our vehicle will go straight or turn which predict that vehicle will take turn. The individual plotting of both features is shown below sets which might be straight which means our vehicle will go straight or turn which predict that vehicle will take turn. The individual plotting of both features is shown below

## 4. Experiment

In this section we are going to present our experiment and results. Firstly, we will discuss the experiment performed to recognize the vehicle license plate number. Then, present our experiment of predicting the route of vehicle in crossroad. For license plate number we use tesseract v4 which is an open source software for recognizing text and is based on deep learning LSTM algorithm [24]. For predicting velocity, we build our own dummy datasets. Our experiment is performed on Windows8.1 Pro 64-bit (6.3, build 9600) Intel(R)Core (TM) i3-400-5U @ 1.7GHZ (4 CPU's).

*4.1.* Vehicle license plate recognition

When the vehicle reaches cross road the surveillance camera attached to opposite vehicle come in action and detected the vehicle. After detection, we apply some basics image processing in order to extract the feature of image for recognizing. When detection and feature extraction is completed, we apply deep learning via tesseract to recognize the license plate number of the vehicle as shown below

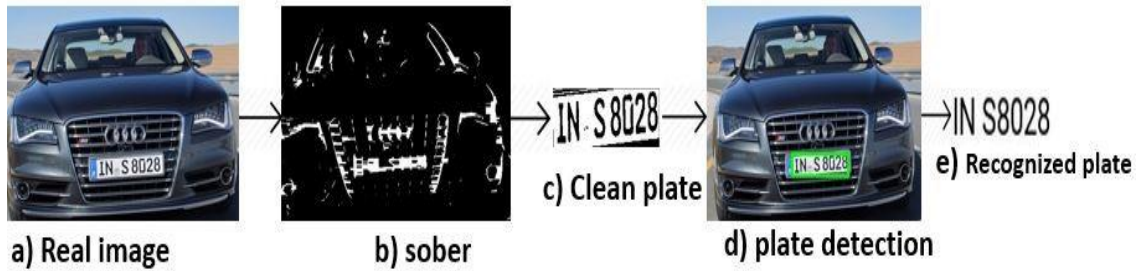

**Figure 4.** Recognition of license plate number

*4.2.* Prediction of route

After recognition of vehicle license plate number, our proposed system checks whether corresponding license number is present in the database repository or not. If not present than terminate the whole prediction process otherwise send the control to the BPU. In this case, vehicle with this license plate number is present in the database that is why send BPU send the first frequency signal from radar gun and get $v1 = 65.5$ and after 3 seconds we send another frequency signal and get velocity $v2 = 65$ after getting both velocities we get the velocities difference $\Delta v = -0.5$.

**Table 3.** Evaluation metrics

|  | precision | Recall | f1-score | support |
|---|---|---|---|---|
| S | 0.98 | 0.96 | 0.96 | 1491 |
| T | 0.98 | 0.96 | 0.96 | 1508 |
| micro avg | 0.98 | 0.96 | 0.96 | 2999 |
| macro avg | 0.98 | 0.96 | 0.96 | 2999 |
| weighted avg | 0.98 | 0.96 | 0.96 | 2999 |

After getting velocities difference we take $\Delta v$ as one of the feature of machine learning algorithm while another feature is previous mobility pattern when we apply machine learning algorithms on these feature and our system predict that vehicle is going to take turn.

## 5. Conclusion

In this paper, due to certain disadvantages of autonomous vehicle we have proposed our own vehicle crossroad route prediction techniques based on machine learning and deep learning. We achieved vehicle route prediction process by using machine learning techniques and in order to achieve this we get facilitation from deep learning by recognition of the vehicles license plate number. In our proposed method, we predict the route of only those vehicles which are registered in our data repository. We predict the route of registered vehicles by using two features one is velocity difference and other is previous mobility patterns then we apply decision tree algorithm on our features and predict whether vehicle will go straight or not. Due to limitation of resources, we use only two features and apply our technique on limited environment. Meanwhile, we use dummy data of previous mobility patterns for our research due non-availability of resources. Our future work will be focused on using more features for accurate prediction of all routes of every vehicle for whom our vehicle mobility might be affected and also use of high-power sensors for increasing the accuracy of our detection and prediction as well.